\documentclass[11pt]{article}

\usepackage[preprint]{acl}

\usepackage{times}
\usepackage{latexsym}
\usepackage{amsthm}
\usepackage{amssymb}
\usepackage{amsmath}
\usepackage{booktabs}
\usepackage{rotating}

\usepackage{tikz}
\usepackage{tikz-cd}
\usetikzlibrary{matrix}
\usetikzlibrary{arrows}
\usetikzlibrary{arrows.meta}
\usetikzlibrary{decorations.pathmorphing,shapes}
\usetikzlibrary{decorations.markings}

\tikzset{spanmap/.style={
            decoration={markings,
            mark= at position 0.5 with {
                  \node[transform shape] (tempnode) {$|$};
                  }
              },
              postaction={decorate}
}
}

\tikzset{doublespanmap/.style={
            decoration={markings,
            mark= at position 0.5 with {
                  \node[transform shape] (tempnode) {$||$};
                  }
              },
              postaction={decorate}
}}

\usepackage[T1]{fontenc}

\usepackage[utf8]{inputenc}

\usepackage{microtype}

\usepackage{inconsolata}

\usepackage{graphicx}

\newcommand{\StoneBraKet}[2]{\langle \, #1 \, | \, #2 \, \rangle}
\newcommand{\subst}[1]{[#1]}
\newcommand{\Eqn}[1]{\mathsf{Eqn}{[#1]}}

\newcommand{\Eqns}{\mathsf{Eqns}}
\newcommand{\Mags}{\mathsf{Mags}}

\newcommand{\ourparagraph}[1]{\medbreak\noindent\textbf{#1}}

\definecolor{littleblue}{rgb}{.2,.2,.6}
\newcommand{\littlebleue}[1]{\textcolor{littleblue}{#1}}

\newcommand{\mathfont}[1]{\littlebleue{#1}}

\newcommand{\ImplicationOrder}{\Rightarrow}
\newcommand{\ImplicationOneStep}{\rightsquigarrow}

\newcommand{\symmetrygroup}[1]{S_{#1}}
\newcommand{\FeatureSpace}[1]{\mathbb{H}_{#1}}

\newcommand{\multiset}[1]{\{\hspace{-.2em}|#1|\hspace{-.2em}\}}

\newcommand{\modulo}[1]{\hspace{.2em}[#1]\hspace{.2em}}

\newcommand{\expectationof}[1]{E[\,#1\,]}
\newcommand{\varianceof}[1]{V[\,#1\,]}

\newcommand{\action}{\circledast}

\title{
The latent space of equational theories}

\author{Luis Berlioz \\
Universidad Nacional \\
Aut{\'o}noma de Honduras \\
  \texttt{luis.berlioz@unah.edu.hm} \\\And
  Paul-Andr{\'e} Melli{\`e}s \\
 Universit{\'e} Paris Cit{\'e} \\
CNRS, INRIA \\
  \texttt{mellies@irif.fr} \\}

\begin{document}
\maketitle
\begin{abstract}
Building on the collaborative
\href{https://teorth.github.io/equational_theories}{Equational Theories}
project initiated by Terence~Tao fifteen months ago,
and combining it with ideas coming from
machine learning and finite model theory,
we construct a \emph{latent space of equational theories}
where each equational theory
is located at a specific location,
determined by its statistical behavior
with respect to
a large sample of finite magmas.
This experiment
enables us to observe for the first time
how reasoning flows
and produces surprisingly oriented 
and well-structured chains
of logical implications in the latent space of equational theories.
%
%
%
\end{abstract}

\begin{small}
\begin{quotation} 
\emph{Two things fill the mind with ever new and increasing admiration and awe, the more often and steadily we reflect upon them: the starry heavens above me and the moral law within~me.}
\\
\vspace{-.8em}
\\
Immanuel Kant, Critique of Practical Reason.
\end{quotation}
\end{small}

\section{Introduction}
This work draws on the recent collaborative project Equational Theories 
(\href{https://teorth.github.io/equational_theories}{ET project})
initiated by Terence Tao about fifteen months ago (Sept.~2024).
In this innovative mathematical experiment,
a large community of thirty-three authors
exchanging on the Lean Zulip
were able to describe \emph{exhaustively} in only a few weeks
the full implication preorder~$\mathfont{(G,\ImplicationOrder)}$
between 4694 basic equational theories on magmas,
see \cite{Bolan2024equational} for details.

Taking inspiration from the \href{https://teorth.github.io/equational_theories}{ET project}
and using the concept of \emph{Stone pairing} 
originating from finite model theory~\cite{Nesetril-Mendez2020},
we experiment
with the idea that mathematical concepts
(in this case, equational theories) can be located 
as \emph{vertices in a latent space},
and then observed and mapped
in the same way as constellations of stars in the sky.

%
%
%
%
%
Once the latent space of equational theories has been defined and constructed,
we derive an \emph{implication graph}~$\mathfont{(G,\ImplicationOneStep)}$
from the full implication preorder~$\mathfont{(G,\ImplicationOrder)}$
made available online by the \href{https://teorth.github.io/equational_theories}{ET project}.
%
This 
allows us to observe
%
that \emph{reasoning flows in a surprisingly well-organized and oriented way}
in the latent space of equational theories.
%
%
%
%
%
%

Despite its simple technical and conceptual design, 
this empirical study at the crossroad of mathematical logic and machine learning
reveals the existence of a rich and well-organised
\emph{latent space of equational theories
and reasoning flows} at the heart of universal algebra.
%
We believe that the discovery
of this emerging piece of mathematical landscape
and the exploration of its remarkable geometric and statistical properties
provide a nice illustration of the benefits of developing
an experimental approach to logic.

\ourparagraph{Plan of the paper.}
After recalling in \S\ref{section/equational-theories-project}
the purpose of the
\href{https://teorth.github.io/equational_theories}{ET project},
we explain in~\S\ref{section/equational-theories-and-implications}
how we extract the graph~$\mathfont{(G,\ImplicationOneStep)}$
of atomic steps from the implication preorder $\mathfont{(G,\ImplicationOrder)}$.
We then define in~\S\ref{section/Stone-pairing} the notion of Stone pairing
from which we derive our feature space and then our latent space of equational theories
in~\S\ref{section/feature-space-of-equational-theories}
and~\S\ref{section/latent-space-of-equational-theories}.
A number of empirical observations on this latent space 
and on the flow of implications are made in~\S\ref{section/latent-space-of-equational-theories}, in~\S\ref{section/clustering-or-provably-equivalent-theories} and \S\ref{section/implication-flows}. 
We describe related works and conclude in 
\S\ref{section/related-works} and \S\ref{section/conclusion}.
We also describe the limitations of our work in \S\ref{section/limitations}.

%

\section{The Equational Theories (ET) project}
\label{section/equational-theories-project}
The purpose of the \href{https://teorth.github.io/equational_theories}{ET project}~\cite{Bolan2024equational}
was to describe entirely
the implication order~$\mathfont{(G,\ImplicationOrder)}$ between 
the 4694 equational theories 
defined with \emph{four instances at most} of the binary connective~$\diamond$
on a given magma $(A,\diamond)$.
This includes fundamental equational theories
such as the \emph{associativity law}
(numbered as $\Eqn{4512}$ in the project)
which involves three variables $x,y,z$
and four instances of the connective:
\vspace{-.1em}
\begin{center}
$\forall x,y,z\in A,
\quad x \diamond ( y \diamond z ) \, = \,
( x \diamond y ) \diamond z$
\end{center}
\vspace{-.1em}
the \emph{commutativity law} ($\Eqn{43}$)
with two variables $x,y$
and two instances of the connective:
\vspace{-.1em}
\begin{center}
$\forall x,y\in A,
\quad x \diamond y \, = \,
y \diamond x$
\end{center}
\vspace{-.1em}
as well as the \emph{idempotence law} ($\Eqn{3}$)
with one variable $x$
and one instance of the connective:
\vspace{-.1em}
\begin{center}
$\forall x\in A,
\quad x \, = \, x \diamond x.$
\end{center}
\vspace{-.1em}
One methodological benefit of the exhaustive approach 
advocated by Tao is that other more exotic and largely unexplored
equational theories suddenly appear as meaningful.
An example is provided by the \emph{Obelix law} ($\Eqn{1491}$)
with four instances of the connective:
\vspace{-.1em}
\begin{center}
$\forall x,y\in A, \quad
x = (y \diamond x) \diamond (y \diamond (y \diamond x))$
\end{center}
\vspace{-.1em}
and the \emph{Asterix law} ($\Eqn{65}$)
with three instances of the connective:
\vspace{-.1em}
\begin{center}
$\forall x, y\in A,
\quad
x = (y \diamond (x \diamond (y \diamond x)))$
\end{center}
\vspace{-.1em}
Indeed, the Asterix law implies 
the Obelix law  for all \emph{finite magmas}
although it is possible to find an \emph{infinite magma}~$(A,\diamond)$
which satisfies the Asterix law without satisfying the Obelix law:
\vspace{-.1em}
\begin{center}
$(A,\diamond)\vDash \Eqn{65} \quad \mbox{and} \quad (A,\diamond)\not\vDash\Eqn{1491}.$
\end{center}
\vspace{-.1em}
where the symbol~$\vDash$ expresses the fact that the magma~$(A,\diamond)$
satisfies a given equational law.
This establishes that there is no proof that $\Eqn{65}$
implies $\Eqn{1491}$, see~\cite{Bolan2024equational} for details.

\section{First step: building the graph~$\mathfont{(G,\ImplicationOneStep)}$}
\label{section/equational-theories-and-implications}

\subsection{The implication preorder $\mathfont{(G,\ImplicationOrder)}$}
We start from the the implication preorder produced by the ET project,
which has $\mathbf{4694}$ elements, one for each equational theory considered in the project.
This means that the number of possible implications between equations is
\vspace{-.1em}
\begin{center}
$4\,694\times 4\,694 = 22\,033\,636$
\end{center}
\vspace{-.1em}
%
It was established by the members of the ET project 
that the actual number of implications
is
\vspace{-.1em}
\begin{center}
$\mathbf{8\,178\,279} \,\, \approx \,\, 
0.37 \times 22\,033\,636$
\end{center}
\vspace{-.1em}
An implication between equational theories
\vspace{-.1em}
\begin{center}
$\Eqn{j} \Rightarrow \Eqn{k}$
\end{center}
\vspace{-.1em}
is called \emph{reversible} when the converse implication 
\vspace{-.1em}
\begin{center}
$\Eqn{k} \Rightarrow \Eqn{j}$
\end{center}
\vspace{-.1em}
holds.
We say in that case that the two equational theories 
are \emph{provably equivalent}
and write
\vspace{-.1em}
\begin{center}
$
\Eqn{j} \sim \Eqn{k}.
$
\end{center}
\vspace{-.1em}
%
%
\noindent
This defines an equivalence relation~$\sim$ which partitions the set of $4\,694$ vertices 
into equivalence classes which we call \emph{reversible cliques}.
Among the implications of the preorder $\mathfont{(G,\ImplicationOrder)}$,
there are 
\vspace{-.1em}
\begin{center}
$
\mathbf{5\,702\,669} \,\, \approx \,\, 0.70 \times 8\,178\,279 
$
\end{center}
\vspace{-.1em}
strict implications and
\vspace{-.1em}
\begin{center}
$
\mathbf{2\,475\,610} \,\, \approx \,\, 0.30 \times 8\,178\,279 
$
\end{center}
\vspace{-.1em}
reversible implications (including the $4\,694$ self-references).

\subsection{The implication graph $\mathfont{(G,\ImplicationOneStep)}$}
Starting from the implication preorder~${(G,\ImplicationOrder)}$
provided by the ET project, we construct a directed 
graph~$(G,\ImplicationOneStep)$ 
of \emph{atomic steps}.
A strict implication
\vspace{-.3em}
\begin{center}
$\Eqn{j}\ImplicationOrder\Eqn{k}$
\end{center}
\vspace{-.1em}
is called \emph{atomic} when it cannot be decomposed,
in the sense that, for all equational theories
$\Eqn{\ell}$,
\vspace{-.1em}
\begin{center}
$\Eqn{j}\ImplicationOrder\Eqn{\ell}
\quad \mbox{and} \quad
\Eqn{\ell}\ImplicationOrder\Eqn{k}$
\end{center}
\vspace{-.1em}
implies that
\vspace{-.1em}
\begin{center}
$
\Eqn{j}\sim\Eqn{\ell}
\quad \mbox{or} \quad
\Eqn{\ell}\sim\Eqn{k}.
$
\end{center}
\vspace{-.1em}
%
We write in that case:
\vspace{-.1em}
\begin{center}
$\Eqn{j}\ImplicationOneStep\Eqn{k}$
\end{center}
\vspace{-.1em}
Note that a strict implication
\vspace{-.1em}
\begin{center}
$\Eqn{j}\ImplicationOrder\Eqn{k}$
\end{center}
\vspace{-.1em}
holds if and only if there exists a path
\vspace{-.1em}
\begin{center}
$
\Eqn{j} \ImplicationOneStep \cdots \ImplicationOneStep \Eqn{k}
$
\end{center}
\vspace{-.1em}
of atomic steps.
It is also worth observing that the notion of atomic step
is not intrinsic, and depends
on the set of equational theories
considered in the~ET project.
The number of atomic steps, or arrows, in the graph $(G,\ImplicationOneStep)$ is
\vspace{-.1em}
\begin{center}
$
\mathbf{1\,052\,209} \,\, \approx \,\, 0.18 \times 5\,702\,669 
$
\end{center}
\vspace{-.1em}

\subsection{The implication graph $\mathfont{(G,\ImplicationOneStep)}$ modulo}
One remarkable property of atomic steps
is that they are closed under reversible implications,
in the sense that every atomic step
$\Eqn{j} \ImplicationOneStep \Eqn{k}$
and provably equivalent equational theories
with the source and target
\vspace{-.1em}
\begin{center}
$\Eqn{j'}\sim\Eqn{j} \quad\quad \Eqn{k}\sim\Eqn{k'}$
\end{center}
\vspace{-.1em}
induces another atomic step
$\Eqn{j'} \ImplicationOneStep \Eqn{k'}$.
For that reason, we declare that two atomic steps
\vspace{-.1em}
\begin{center}$
\Eqn{j} \ImplicationOneStep \Eqn{k}
\quad\quad
\Eqn{j'} \ImplicationOneStep \Eqn{k'}
$
\end{center}
\vspace{-.1em}
are equivalent modulo~$\sim$ when
$\Eqn{j}\sim\Eqn{j'}$ and $\Eqn{k} \sim \Eqn{k'}$.
%
%
We obtain in this way a directed acyclic graph noted $(G,\ImplicationOneStep)/\!\sim$
\vspace{-.1em}
\begin{center}
with $\mathbf{1415}$ vertices 
and $\mathbf{4824}$ arrows
\end{center}
\vspace{-.1em}
defined as equivalence classes of vertices and arrows of $(G,\ImplicationOneStep)$
modulo~$\sim$.
%



\section{Stone pairing}\label{section/Stone-pairing}
We find useful to recall the notion of \emph{Stone pairing} introduced
by~\cite{Nesetril-Mendez2012, Nesetril-Mendez2020}
in the field of finite model theory,
see also \cite{GehrkeJaklReggio2022}
for a discussion.
We will see in~\S\ref{section/feature-space-of-equational-theories}
and~\S\ref{section/latent-space-of-equational-theories} that
our definition of the feature space and of the latent space 
of equational theories relies on this \emph{probabilistic}
variant of validity
of a first-order formula~$\varphi$ in a model~$A$.
%
%
%
Here, we focus on the \emph{first-order theory of magmas}
defined by the signature~$\{\diamond:2\}$ 
for the binary function~$\diamond$.
Recall that a magma~$(A,\diamond)$ is
defined as a set~$A$ equipped with a binary operation 
\vspace{-.1em}
\begin{center}
$\diamond \,\, : \,\, A\times A\longrightarrow A$.
\end{center}
\vspace{-.1em}
A typical first-order predicate~$\varphi$ 
states the existence of a neutral element:
\vspace{-.1em}
\begin{center}
$\exists e. \forall x. \quad 
\big(\, x \diamond e = x \,\, \wedge \,\, e\diamond x = x\, \big)$
\end{center}
\vspace{-.1em}
or the existence of a right residual for any pair of elements~$x,z$ of the magma:
\vspace{-.1em}
\begin{center}
$\forall x. \forall z. \exists y. \quad x \diamond y = z.$
\end{center}
\vspace{-.1em}
All equations between magmas can be expressed in that way.
%
Typically, $\Eqn{3721}$ of the equational theory project can be written as
the formula
\vspace{-.1em}
\begin{center}
$\varphi(x,y) \quad \equiv \quad 
x \diamond y  \,\, = \,\, (x \diamond y)\diamond(x \diamond y).$
\end{center}
\vspace{-.1em}
with two free variables $x,y$.
%
%
Now, given a finite magma~$(A,\diamond)$
and a first-order formula
\vspace{-.1em}
\begin{center}
$\varphi(x_1,\dots,x_n)$
\end{center}
\vspace{-.1em}
with $n$ free variables~$x_1,\dots,x_n$,
the Stone pairing
\vspace{-.1em}
\begin{center}
$\StoneBraKet{\varphi}{A}\, \in \, [0,1]$
\end{center}
\vspace{-.1em}
is the scalar defined as the probability:
\vspace{-.1em}
\begin{center}
$
\frac{\#\,\big\{ \, (a_1,\dots,a_n)\in A^n \,\, | \,\, 
\varphi(a_1,\dots,a_n) \, \mbox{holds} \, \big\}}{\# A\, ^n}
$
\end{center}
\vspace{-.1em}
that a $n$-tuple of elements
\vspace{-.1em}
\begin{center}
$(a_1,\dots,a_n)\in A^n$
\end{center}
\vspace{-.1em}
satisfies the  $n$-ary formula~$\varphi$.
%
Here we write $\# A$ for the cardinal of a finite set~$A$.
By way of illustration, observe that
\vspace{-.1em}
\begin{center}
$A \,\, \vDash \,\, \forall \mathbf{x}, \varphi (\mathbf{x})
\quad \iff \quad
\StoneBraKet{\,\varphi(\mathbf{x})\,}{\,A\,}\,\, = \,\, 1$
\end{center}
\vspace{-.1em}
where we write $A\vDash \psi$ when a first-order formula~$\psi$
is satisfied by a given magma~$A$,
and use the notation $\mathbf{x}=x_1,\dots,x_n$ for the sequence of free variables
of the formula~$\varphi$.
From this follows in particular that
\vspace{-.1em}
\begin{center}
$
\StoneBraKet{\,x=x\,}{\,A\,}\,\, = \,\, 1
$
\end{center}
\vspace{-.1em}
for every finite magma~$A$.
Fractional values different from $0$ and $1$ are the most common
in practice.
For instance, note that
\vspace{-.1em}
\begin{center}
$
\StoneBraKet{\,x=y\,}{\,A\,}\,\, = \,\, \frac{1}{\,\# A\,}
$
\end{center}
\vspace{-.1em}
for every finite magma~$A$, since 
there is a probability of one divided by the cardinal of~$A$ that
two elements $a_1, a_2\in A$ happen to be equal.
%

\section{The feature space of equational theories}\label{section/feature-space-of-equational-theories}
In this section, we explain how we use the probabilistic concept
of Stone pairing in order to construct
our latent space of equational theories.

\subsection{The generation process}
We proceed in the following way.
We start by generating a large number $n=\#\Mags$ 
of finite magmas~$A_1,\dots,A_n$ of a fixed size~$N$.
For simplicity, we define them 
with the same underlying set~$\{0,\dots,N-1\}$.
Each magma~$A_{\ell}$ for $1\leq\ell\leq n$
is thus generated as a two-dimensional array of size $N\times N$, 
where each entry is a randomly chosen value between 0 and $N-1$. 
This two-dimensional array describes the multiplication table 
of the magma~$A_{\ell}=\{0,\dots,N-1\}$ 
and thus completely characterizes it.
We typically work with a sample of $n=1000$ magmas of size~$N$
between $4$ and $16$.
%
These numbers~$n=1000$ and~$4\leq N\leq 16$
appear reasonable from a computational point of view
and sufficient for our purposes.

\subsection{The feature hypercube $\FeatureSpace{n}=[0,1]^n$}
\begin{figure}
\centering
\includegraphics[width=0.5\textwidth]{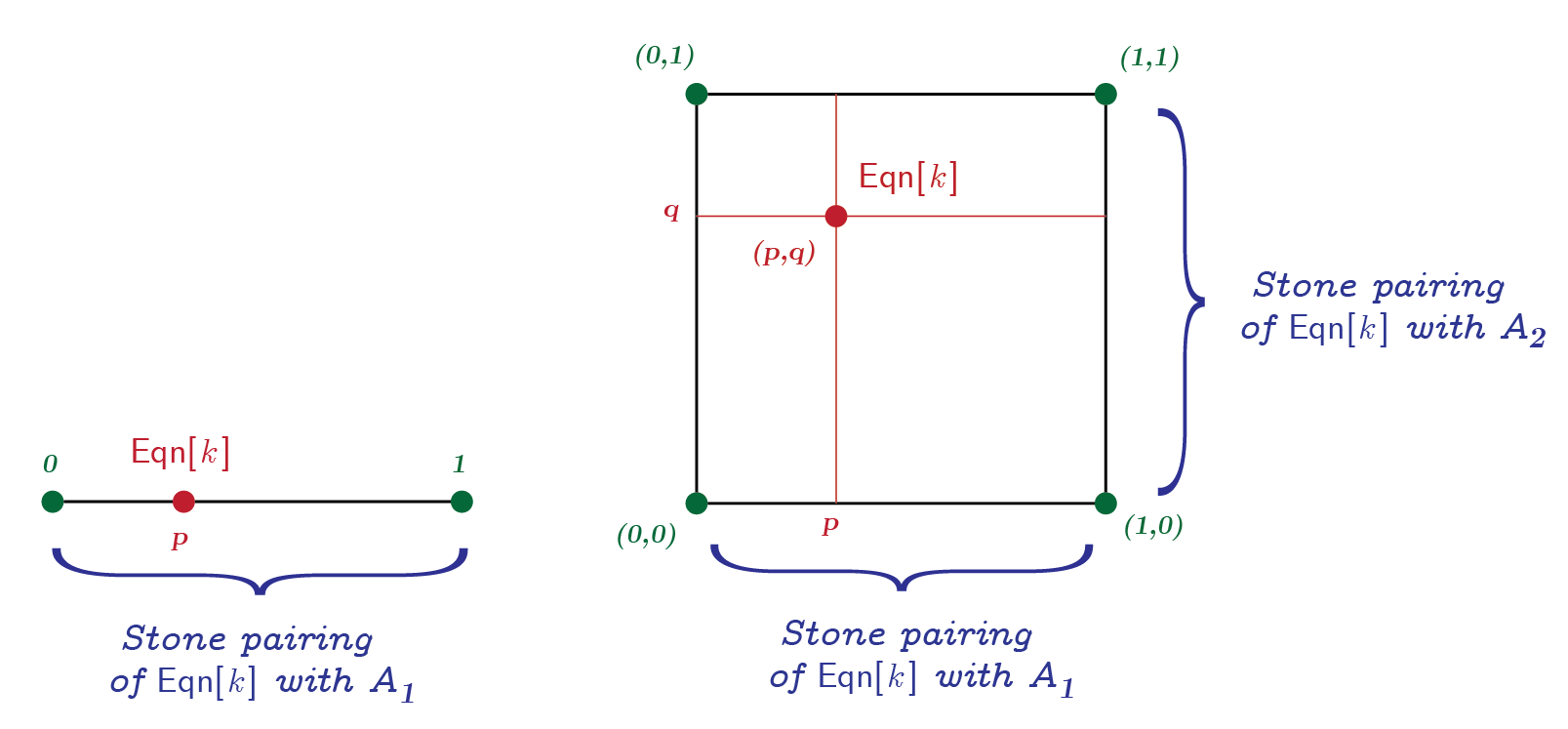}
\caption{How each equation $\Eqn{k}$ is assigned a position in the hypercube $\FeatureSpace{n}$ of dimension~$n=\#\Mags$ the number of sampled finite magmas $A_1,\dots,A_n$.}
\label{figure/hypercube}
\end{figure}

%
We define a matrix of size $\#\Eqns\times\#\Mags$ 
\begin{equation}\label{equation/matrix-R}
R\quad = \quad \big(\, p_{k,\ell}\,\big)_{k\in\Eqns,\ell\in\Mags}
\end{equation}
where $\#\Eqns = 4964$ is the number of equations
and $n=\#\Mags=1000$ is the number of sampled magmas of size~$N$.
Each entry~$p_{k,\ell}$ of the matrix~$R$
is defined as the probability
\vspace{-.1em}
\begin{center}
$p_{k,\ell} \quad = \quad \StoneBraKet{\Eqn{k}}{A_{\ell}}$
\end{center}
\vspace{-.1em}
given by the Stone pairing of $\Eqn{k}$ with the finite magma~$A_{\ell}$.
Every equation $\Eqn{k}$ induces a line or vector of $n$ probabilities
\vspace{-.1em}
\begin{center}
$
\big(\,p_{k,1} \,\, , \,\, \dots \,\, , \,\, p_{k,n}\,\big) \,\,\in \,\, [0,1]^n
$
\end{center}
\vspace{-.1em}
which defines an element of the hypercube $\FeatureSpace{n}=[0,1]^{n}$ of dimension~$n=\#\Mags$,
see Fig.~\ref{figure/hypercube} for an illustration.
%

\medbreak

\subsection{Stone spectrum}\label{section/Stone-Spectrum}
\begin{figure}
    \centering
    \includegraphics[width=0.85\linewidth]{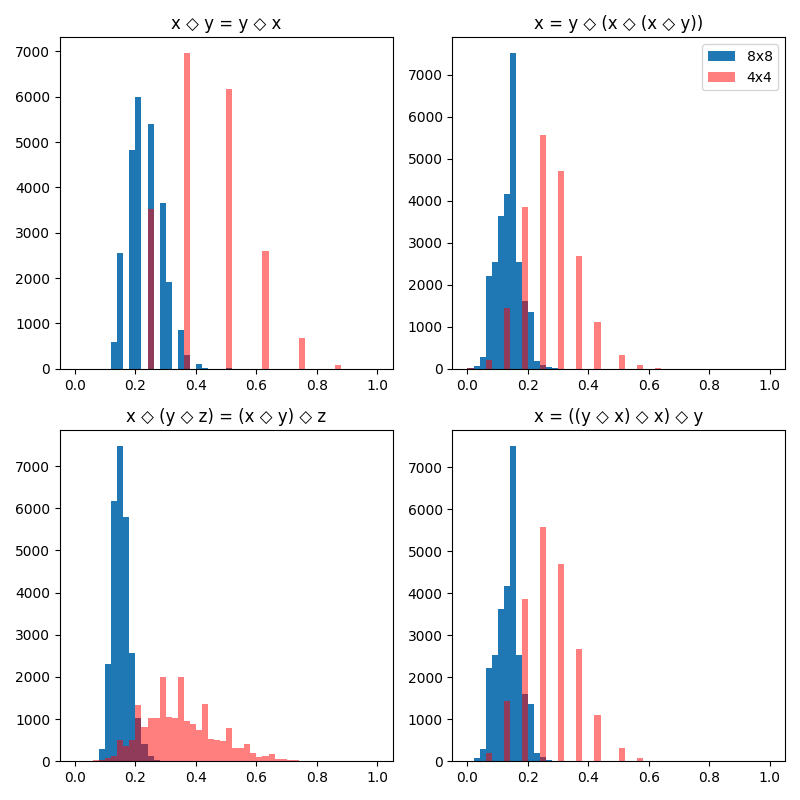}
    \caption{Stone spectra of the commutative law (top-left) and of the associative law (bottom-left) for a sampling of $n=20\,000$ magmas of size 4 (red curve) and of $n=24\, 000$ magmas of size 8 (blue curve). The Stone spectra of the conjugate theories $\Eqn{63}$ and $\Eqn{271}$ are identical (top and bottom, right).}
    \label{figure:stone-spectra}
\end{figure}

The hypercube $\FeatureSpace{n}=[0,1]^{n}$ comes equipped with an action 
\vspace{-.1em}
\begin{center}
$
\begin{tikzcd}[column sep = 1em, row sep = 0em]
\action \quad : \quad \symmetrygroup{n} \times \FeatureSpace{n} \arrow[rr] && \FeatureSpace{n}
\end{tikzcd}
$
\end{center}
\vspace{-.1em}
of the symmetry group~$\symmetrygroup{n}$ 
on the $n$ dimensions (or directions) of the space $\FeatureSpace{n}$,
defined as
\vspace{-.1em}
\begin{center}
$
\sigma \action (p_1,\dots,p_n)
:= (p_{\sigma(1)},\dots,p_{\sigma(n)})
$
\end{center}
\vspace{-.1em}
for $\sigma\in\symmetrygroup{n}$ and $\mathbf{p}=(p_1,\dots,p_n)\in\FeatureSpace{n}$.
Since the sampling order of the magmas~$A_1,\dots, A_n$ does not matter,
it makes sense to consider the position of each equational theory
\vspace{-.1em}
\begin{center}
$
\mathbf{p} \, := \, (p_1,\dots,p_n) \, \in \FeatureSpace{n}
$
\end{center}
\vspace{-.1em}
modulo the action of the symmetry group~$\symmetrygroup{n}$.
One obtains in this way a position in the quotient space $\FeatureSpace{n}/\symmetrygroup{n}$,
defined as the multiset
\vspace{-.1em}
\begin{center}
$\mathbf{p} \modulo{\symmetrygroup{n}} \, := \, \multiset{\, p_1 \, , \, \dots \, , \, p_n\,}$
\end{center}
\vspace{-.1em}
We call this multiset
the \emph{Stone spectrum} of the equational theory.
%
The Stone spectrum can be nicely visualized 
as a finite positive measure
of weight $n$ on the interval $[0,1]$,
defined as the sum
\vspace{-.1em}
\begin{center}
$
\sum_{i=1}^{n} \, \delta_{p_i}
$
\end{center}
\vspace{-.1em}
of the Dirac distributions $\delta_{p_i}$, see Fig.~\ref{figure:stone-spectra} for an illustration.

\medbreak

\subsection{Stone interference spectrum}
The cartesian product of the hypercube $\FeatureSpace{n}=[0,1]^{n}$ with itself 
comes equipped with an action
\vspace{-.1em}
\begin{center}
\begin{tikzcd}[column sep = 1em, row sep = 0em]
{\action} \quad : \quad \symmetrygroup{n} \times \FeatureSpace{n} \times \FeatureSpace{n} \arrow[rr] && \FeatureSpace
\times \FeatureSpace{n}
\end{tikzcd}
\end{center}
\vspace{-.1em}
defined componentwise:
\vspace{-.1em}
\begin{center}
$\sigma\action(\mathbf{p},\mathbf{q}) \,\, := \,\, (\sigma\action\mathbf{p},\sigma\action\mathbf{q})$
\end{center}
\vspace{-.1em}
Instead of looking at the relative positions of a pair of equational theories~$\Eqn{j}$
and~$\Eqn{k}$ in the feature space $\FeatureSpace{n}\times\FeatureSpace{n}$, we may look 
at their relative positions in the quotient space $\FeatureSpace{n}\times\FeatureSpace{n}/\symmetrygroup{n}$.
This is provided by the finite multiset of pairs:
\vspace{-.1em}
\begin{center}
$\mathbf{(p,q)} \modulo{\symmetrygroup{n}} \, := \, \multiset{\, (p_1,q_1) \, , \, \dots \, , \, (p_n,q_n)\,}$
\end{center}
\vspace{-.1em}
which we call \emph{Stone interference spectrum}.
In the same way as the Stone spectrum of an equational theory,
this interference spectrum of two equational theories
can be described as the finite measure on the square $[0,1]^2$
defined as the sum
\vspace{-.1em}
\begin{center}
$\sum_{i=1}^{n}\, \sum_{j=1}^{n} \, \delta_{(p_i,q_j)}$
\end{center}
\vspace{-.1em}
where $\delta_{(p,q)}$ denotes the Dirac distribution of $(p,q)\in[0,1]^2$.
This representation enables one to visualize
the Stone interference spectrum in an instructive way, 
see Fig.~\ref{figure:stone-spectra} for two illustrations.

\begin{figure}
\centering
\includegraphics[width=0.5\textwidth]{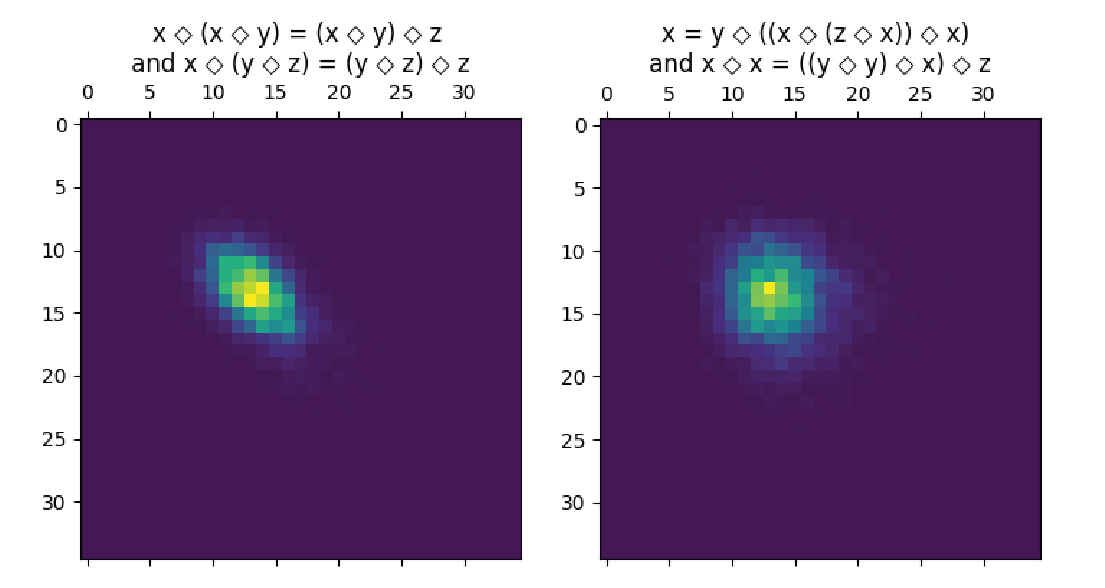}
\caption{The Stone interference spectrum of $\Eqn{4400}$ and $\Eqn{4533}$ detects a degree of statistical dependence between the two conjugate equational theories (left) while the Stone interference spectrum of $\Eqn{1092}$ and $\Eqn{4092}$ shows full statistical independence between the two equational theories (right).}
\end{figure}


\section{The latent space of equational theories}
\label{section/latent-space-of-equational-theories}
%

\subsection{Building the latent space
using a PCA dimensionality reduction}
We apply a principal component analysis (PCA) method in order 
to derive a 3-dimensional representation of the $4694$ equational theories
from thr positions in the $n$-dimensional feature space~$\FeatureSpace{n}$.
We thus start by computing the center of gravity $\mu\in\FeatureSpace{n}$
of the $4694$ vertices in~$\FeatureSpace{n}$.
We subtract $\mu$ from each column of the matrix~$R$
defined in \eqref{equation/matrix-R},
in order to obtain a matrix $X=R-\mu$ with zero-mean rows.
The PCA method then diagonalizes the covariance matrix $C_X = \frac{1}{n}\,XX^T$,
see \cite{shlens2014tutorial, hanchi2025geometric} for details.
%
%

\begin{figure}
    \centering
    \includegraphics[width=1\linewidth]{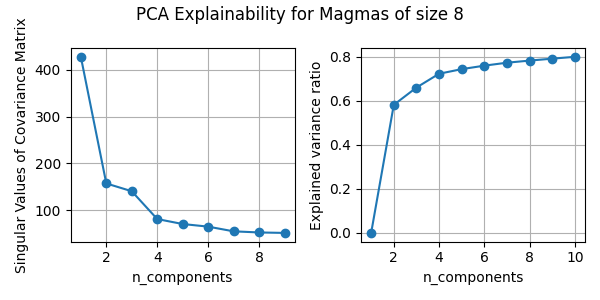}
    \caption{PCA Explainability statistics for a random sample of $n=20\,000$ magmas of size 8. The values of the first ten singular values (left) and the explained variance ratio (right) for the first ten principal components.}
    \label{fig:explain_ratio}
\end{figure}

We keep the first three principal components
or dimensions (noted~$X, Y, Z$) of the row-space which 
we call the \emph{latent space of equational theories}. 
This choice of picking the three first dimensions
of the PCA is justified by our visualization purposes
as well as by the fact that
a majority of variability in the data is captured by 
the first three principal components, as shown in Fig.~\ref{fig:explain_ratio}.

\subsection{The latent space
is good at detecting signatures of equational theories}
\label{subsection/good-at-detecting-signatures}
\begin{figure}
\centering
\includegraphics[width=.99\linewidth]{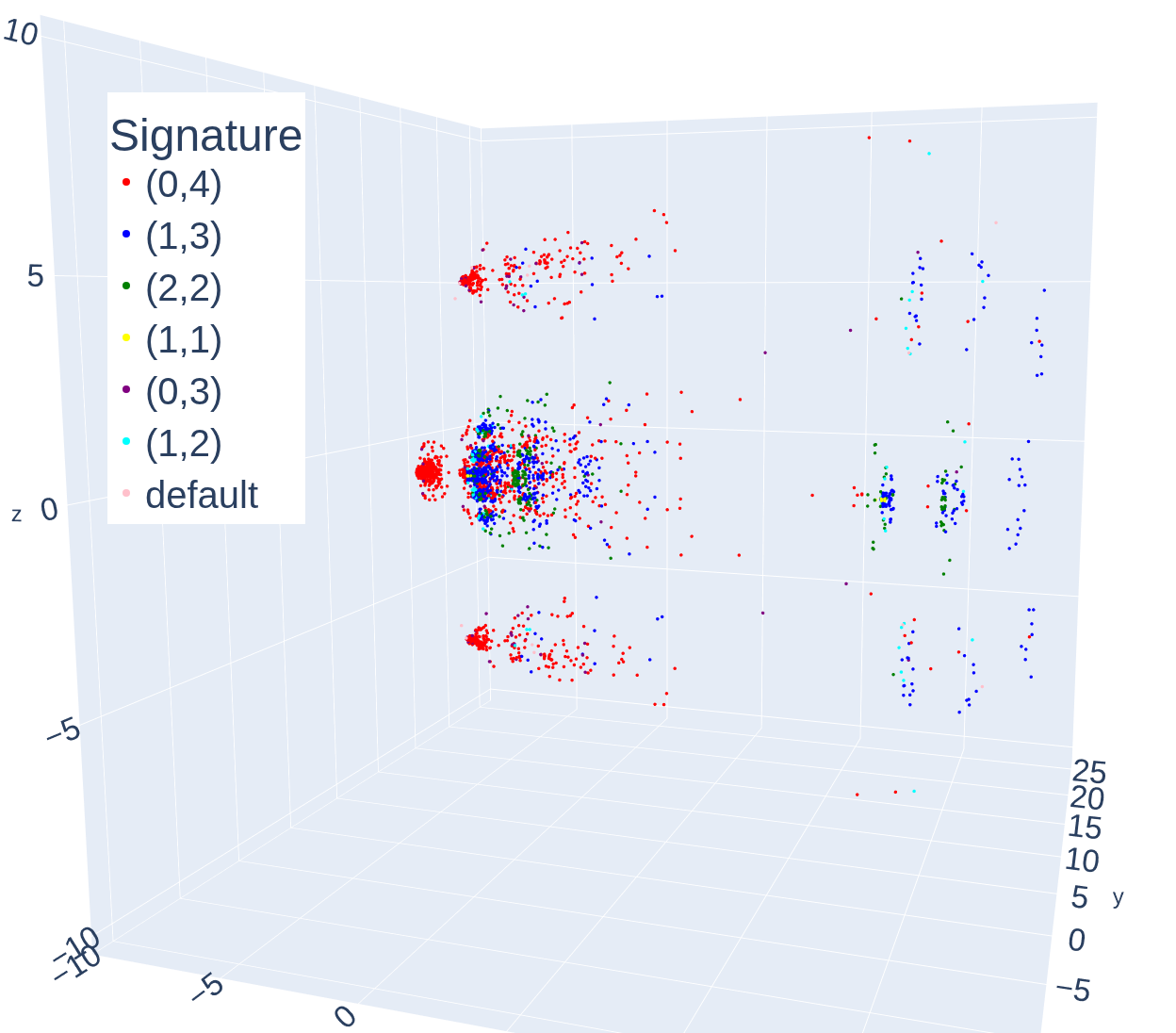}
\caption{Latent space of equational theories where every theory is colored according to its signature~$(a,b)$, where $a$ and $b$ are the numbers of diamond operations on the lefthand and righthand side of the theory.}
\label{figure:latent-space-with-signatures}
\end{figure}

Exploring the latent space and looking at each position of the $4964$ equational theories, we observe that the latent space is surprisingly good at identifying and grouping together the equational theories $\Eqn{k}$ with the same signature, see Fig.~\ref{figure:latent-space-with-signatures}.

Here, by signature of an equational theory, we mean the ordered pair $(a,b)$ describing the numbers $a$ and $b$ of diamonds on the left-hand side and right-hand side of the equation, respectively.
Typically, the Tarski's law~$\Eqn{542}$ 
\vspace{-.1em}
\begin{center}
$x \, = \, (z \diamond (x \diamond (y \diamond z)))$
\end{center}
\vspace{-.1em}
has signature $(0,3)$.
We indicate how the $4694$ equational theories are partitioned according to their signature, and how they are colored in the latent space depicted in Fig.~\ref{figure:latent-space-with-signatures}.

\begin{center}
\begin{footnotesize}
\fbox{
\begin{tabular}{cccc}
$2842$ & theories have signature & $(0,4)$ & [red]
\\
$1015$ & theories have signature & $(1,3)$ & [blue]
\\
$427$ & theories have signature & $(2,2)$ & [green]
\\
$260$ & theories have signature & $(0,3)$ & [purple]
\\
$104$ & theories have signature & $(1,2)$ & [cyan]
\\
$30$ & theories have signature & $(0,2)$ & [pink]
\\
$9$ & theories have signature & $(1,1)$ & [yellow]
\\
$5$ & theories have signature & $(0,1)$ & [pink]
\\
$2$ & theories have signature & $(0,0)$ & [pink]
\end{tabular}}
\end{footnotesize}
\end{center}


%

\subsection{Expectation and variance}\label{section/EV}
The \emph{expectation} and \emph{variance} of an equational theory~$\Eqn{k}$
are defined as the expectation and variance of the associated 
sequence of Stone pairings:
\begin{center}
\begin{tabular}{c}
$
\expectationof{\Eqn{k}} \, = \, \frac{1}{n}\sum_{i=1}^{n} p_{i}
$
\\
\vspace{-1em}
\\
$
\varianceof{\Eqn{k}} \, = \, \frac{1}{n}\sum_{i=1}^{n} \Big( \, p_i - \expectationof{\Eqn{k}} \Big)^{2}
$
\end{tabular}
\end{center}
where $p_i=\StoneBraKet{\Eqn{k}}{A_{i}}$.
Note that the expectation and variance do not depend on the sampling order 
of the finite magmas $A_1,\dots,A_{n}$
and can be thus computed from the Stone spectrum of~$\Eqn{k}$.

\subsection{The $X$-axis is strongly correlated to expectation}
\label{subsection-X-axis}
The first principal component~$X$ of a theory~$\Eqn{k}$
is strongly correlated to its expectation~$\expectationof{\Eqn{k}}$,
as shown by the linear regression graph, see Fig.~\ref{fig:mean-pca} in Appendix.
As a consequence,
the theory with maximum $X$-value is $\Eqn{0}$
\vspace{-.1em}
\begin{center}
$
x\, = \, x
$
\end{center}
\vspace{-.1em}
with expectation exactly~$1$, followed by $\Eqn{3277}$ and its conjugate $\Eqn{4067}$
\vspace{-.1em}
\begin{center}
\begin{tabular}{c}
$
x\diamond x \, = \, ((x\diamond x)\diamond y)\diamond y
$
\\
$
x\diamond x \, = \, y\diamond (y\diamond (x\diamond x))
$
\end{tabular}
\end{center}
\vspace{-.1em}
whose expectation in the sample of finite magmas of size $8$
we took is around $.46$.
%
On the other side of the latent space, one of the equational theories 
with smallest~$X$-value is $\Eqn{1}$
\vspace{-.1em}
\begin{center}
$
x\, = \, y
$
\end{center}
\vspace{-.1em}
with expectation exactly $\frac{1}{8}=.125$ in any finite magma of size $N=8$
used in the sample.

\subsection{The $Y$-axis is correlated to variance}
\label{subsection-Y-axis}
The principal component~$Y$ of an equational theory~$\Eqn{k}$
is strongly correlated to its variance~$\varianceof{\Eqn{k}}$,
as shown by the linear regression, see Fig.~\ref{fig:std-pca} in Appendix.
The equational theory with minimum variance and $Y$-value is $\Eqn{4275}$ and conjugate $\Eqn{4590}$:
\vspace{-.1em}
\begin{center}
\begin{tabular}{c}
$
x\diamond (x\diamond x) \, = \, y\diamond (y\diamond y)
$
\\
$
(x\diamond x)\diamond x \, = \, (y\diamond y)\diamond y
$
\end{tabular}
\end{center}
\vspace{-.1em}
The equational theory with maximum variance and $Y$-value is $\Eqn{3658}$ 
which is self-conjugate:
\vspace{-.1em}
\begin{center}
$
x\diamond x \, = \, (x\diamond x) \diamond (x\diamond x)
$
\end{center}
\vspace{-.1em}

\subsection{The $Z$-axis detects conjugacy}
One immediate and striking observation on the latent space 
is the $Z$-value produces a reflection
symmetry between conjugate theories,
around the plane $Z=0$.
We noticed this emerging reflection symmetry between conjugate theories
in our first experiments, and improved this reflection symmetry
by selecting a perfectly symmetric sample of finite magmas.
For reference, the equational theory with maximum $Z$-value 
is $\Eqn{1022}$ 
\vspace{-.1em}
\begin{center}
$
x \, = \, x\diamond ((x\diamond (x\diamond y))\diamond y)
$
\end{center}
\vspace{-.1em}
closely followed by $\Eqn{428}$
\vspace{-.1em}
\begin{center}
$
x\, = \,  x\diamond (y\diamond (x\diamond (y\diamond x)))
$
\end{center}
\vspace{-.1em}
By reflection symmetry, the theory with minimum $Z$-value 
is the conjugate $\Eqn{2529}$:
\vspace{-.1em}
\begin{center}
$
x \, = \, (y\diamond ((y\diamond x)\diamond x)\diamond x
$
\end{center}
\vspace{-.1em}
closely followed by the conjugate $\Eqn{3067}$:
\vspace{-.1em}
\begin{center}
$
x\, = \,  (((x\diamond y)\diamond x)\diamond y)\diamond x
$
\end{center}
\vspace{-.1em}
Accordingly,
there are 84 self-conjugate equational theories
and they all appear on the plane $Z=0$.
This ability of the $Z$-component to detect conjugacy is remarkable
since two conjugate theories~$\Eqn{j}$ and~$\Eqn{k}$
have the same Stone spectrum, and thus the same expectation and variance,
see the definition in~\S\ref{section/Stone-Spectrum} and discussion 
in~\S\ref{section/EV}.
%

%
%


\section{Clustering of provably equivalent theories in the latent space}
\label{section/clustering-or-provably-equivalent-theories}
%
%
%
Now that we have constructed the latent space of equational theories,
we can embed the implication graph $(G,\ImplicationOneStep)$
to visualize what it looks like in three dimensions.
For every pair of equational theories
related by a proof
    $\Eqn{j}\Longrightarrow\Eqn{k}$
we thus draw an \emph{edge} 
    $\Eqn{j}\longrightarrow\Eqn{k}$
between the vertices associated 
to $\Eqn{j}$ and $\Eqn{k}$
in the latent space.
%
%
We use the same terminology as for implications, and thus distinguish three classes of edges: 
the \emph{reversible} edges $\sim$,
the \emph{atomic} edges $\ImplicationOneStep$ which are strict by definition,
and the \emph{strict} edges $\ImplicationOrder$ which are general and not necessarily atomic.
%
The length of an edge can be computed in several ways.
In this paper, we choose the simplest distance which is the Euclidian distance in the latent space.

At this stage, it is important to notice that two provably equivalent theories $\Eqn{j}$ and $\Eqn{k}$
may induce different Stone pairings within a given finite magma~$(A,\diamond)$.
Indeed, $\Eqn{j}\sim\Eqn{k}$ simply ensures that validity is preserved
\begin{center}
$(A,\diamond)\vDash\Eqn{j} \,\, \iff \,\, (A,\diamond)\vDash\Eqn{k}$
\end{center}
in other words, that the Stone pairing of $\Eqn{j}$ is equal to $1$ 
if and only if the Stone pairing of $\Eqn{k}$ is equal to $1$.
%
On the other hand, we observe empirically that the reversible edges
are on average seven times shorter than strict atomic edges,
and ten times shorter than general stric edges,
as indicated in the table below.



\begin{figure}[h!]
\centering
\includegraphics[width=.99\linewidth]{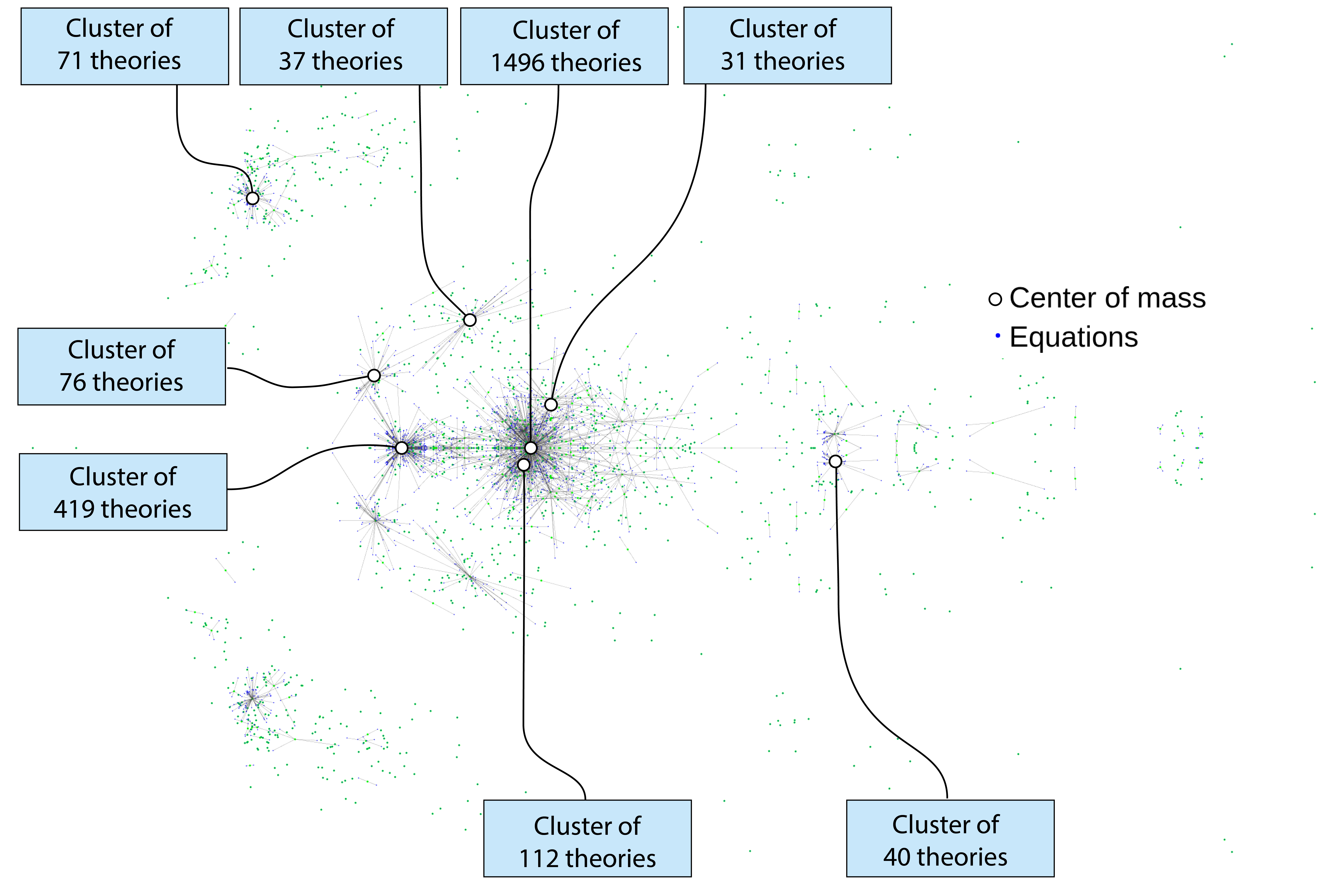}
\caption{Visualizing the center of mass of each provably equivalent clique. Each equational theory is drawn connected to its corresponding center of mass.}
\label{figure/center-of-mass}
\end{figure}

\medbreak
\begin{center}
\begin{small}
    \begin{tabular}{crrr}
    \toprule
    & \textbf{Reversible} & \textbf{Atomic} & \textbf{Strict}\\
    \midrule
count & 2\,470\,916 & 1\,052\,209 &  5\,707\,363\\
mean & 0.69 & 5.29 &  7.17\\
std & 0.65 & 2.22 &  4.30\\
min & 0.00 & 0.00 & 0.00 \\
25\% & 0.00 & 4.14 & 4.17 \\
50\% & 0.63 & 4.69 & 6.14 \\
75\% & 0.96 & 6.33 & 9.22 \\
max & 5.36 & 37.08 & 41.20 \\
\bottomrule
    \end{tabular}
    \end{small}  
\end{center}    
\medbreak
\noindent 
This important difference suggests a degree of clustering of the provably equivalent theories in the latent space.
The hypothesis can be validated
by computing the center of mass of each clique
of provably equivalent theories, and connecting each theory
to the center of mass of its corresponding clique.
One obtains in this way a three-dimensional picture,
see Fig.~\ref{figure/center-of-mass}, where the clusters
of provably equivalent theories are clearly separated.

\begin{figure}
\centering
\includegraphics[width=.99\linewidth]{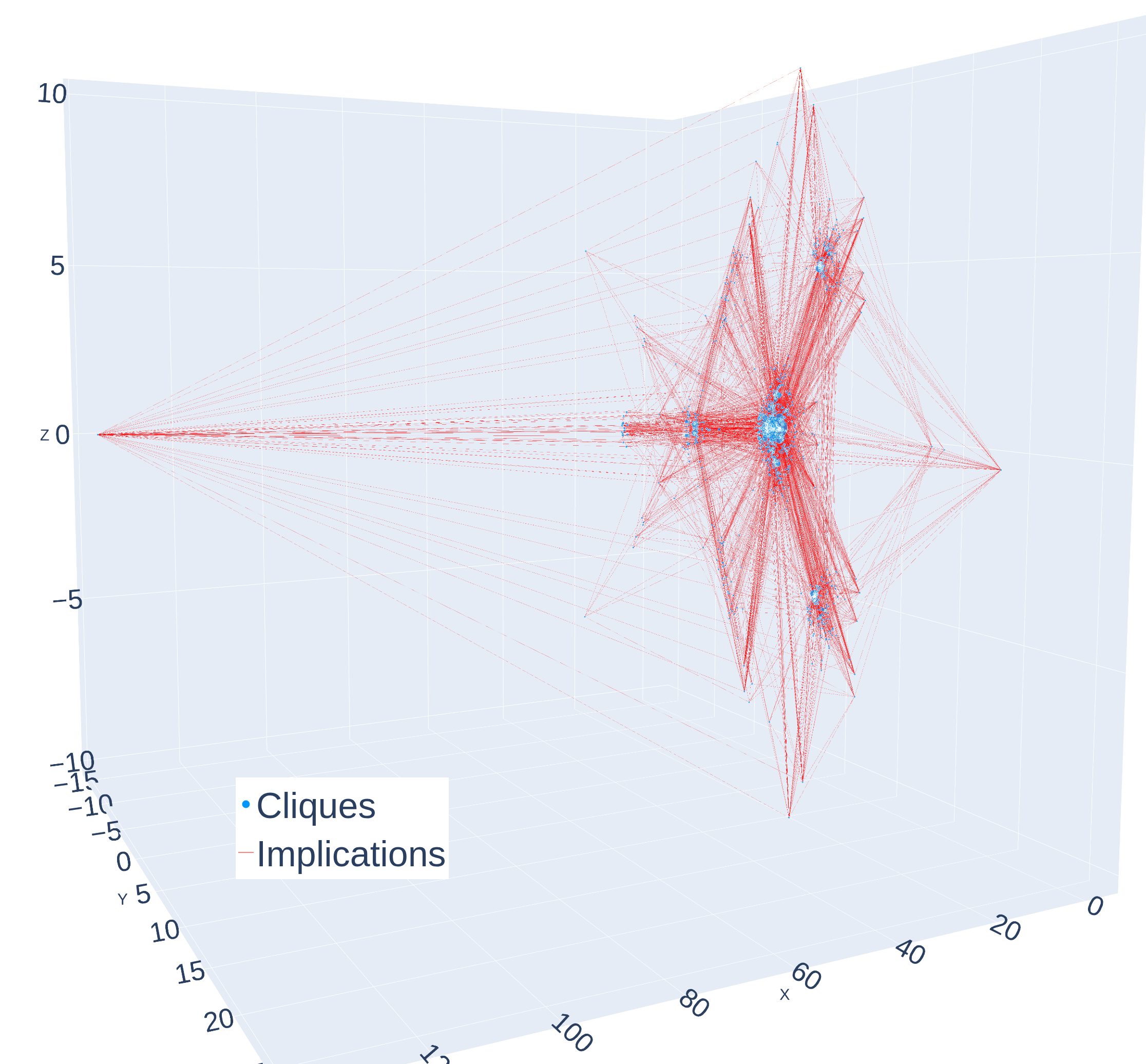}
\caption{A global view of the latent space of equational theories and of the $1415$ vertices and $4824$ edges of the implication graph $(G,\ImplicationOneStep)$ modulo reversibility. The existence of implication flows reveals an emerging similitude between the model-theoretic definition of the latent space and the proof-theoretic definition of the implication graph.}
\label{fig:goldfish}
\end{figure}

\section{Implication flows in the latent space}\label{section/implication-flows}
The clustering of provably equivalent theories in the latent space 
observed in~\S\ref{section/clustering-or-provably-equivalent-theories}
has an important consequence: it enables us 
to derive from the original embedding of $(G,\ImplicationOneStep)$
an embedding of the graph $(G,\ImplicationOneStep)$ modulo~$\sim$.
The idea is to replace each reversible clique modulo~$\sim$ by a single ball
located at the center of mass of the associated cluster.
The number of provably equivalent theories in the clique 
is indicated by the diameter of the ball.
One obtains in that way a graph with 1415 vertices (blue balls of different diameters) 
and 4824 edges (red arrows), see Fig.~\ref{fig:goldfish}.
%

\medbreak

We mention a number of remarkable emerging phenomena 
which appear when one observes the embedding of the graph 
$(G,\ImplicationOneStep)$ modulo~$\sim$.


\subsection{Radial direction of implication flows}
A striking observation on the picture of Fig.~\ref{fig:goldfish}
is the existence of clearly oriented implication flows
across the latent space.
The mainstream direction goes along the $X$-axis 
towards equational theories with higher expectations.
There is also a clear radial orientation along the $Y$ and $Z$-axis
towards theories with higher variance, and with different degrees
of self-conjugacy.
The implication graph looks disorganised close to the source $\Eqn{1}$
which implies all the other theories.
On the other hand, it looks very organised close to the sink $\Eqn{0}$
which is implied by every theory.
One tentative explanation is that more equational theories $\Eqn{k}$
such as \emph{idempotency} or \emph{commutativity} 
can be deduced as consequences of a theory $\Eqn{j}$ closer to the source.
These equational theories can be used in the reasoning
to establish unexpected facts, and produce disorganised edges.
On the other hand, when one gets closer to the sink,
it becomes much more difficult to prove things,
and this could explain why there are so few implication edges
going against the mainstream radial orientation

%
%
%
%

\subsection{Parallelism of implications}
Another striking phenomenon is that many implication edges appear to be similar and parallel in the latent space.
A closer inspection shows that very often, the parallel edges have similar source and target theories.
An illustration is provided by the pair of parallel implications
\vspace{-.1em}
\begin{center}
\begin{small}
$\begin{array}{c}
\Eqn{422}\Longrightarrow \Eqn{2851}
\\
\Eqn{427}\Longrightarrow \Eqn{2861}
\end{array}$
\end{small}
\end{center}
\vspace{-.1em}
where $\Eqn{422}$ and $\Eqn{427}$ are very similar:
\begin{small}
\begin{center}
\begin{tabular}{c}
$\forall x.y.z. \quad x \, = \, x\diamond (x\diamond (y\diamond (z\diamond y)))$
\\
$\forall x.y.z. \quad x \, = \, x\diamond (y\diamond (x\diamond (x\diamond z)))$
\end{tabular}
\end{center}
\end{small}
and for $\Eqn{2851}$ and $\Eqn{2861}$:
\begin{small}
\begin{center}
\begin{tabular}{c}
$\forall x.y. \quad x \, = \, ((x\diamond (x\diamond y))\diamond x)\diamond x$
\\
$\forall x.y. \quad x \, = \, ((x\diamond (y\diamond x))\diamond x)\diamond x$
\end{tabular}
\end{center}
\end{small}
%
We conjecture that this parallelism phenomenon
is related to the Herbrand theorem in formal logic,
see~\cite{Miller87} and the discussion in Appendix.
Note in particular that the witnesses (or expansion trees)
provided by the Herbrand theory generalise the notion of \emph{simple rewrite}
elaborated and collected
in \href{https://teorth.github.io/equational_theories/blueprint/simple-rewrite-chapter.html}{Section 21
of the ET project blueprint}.
In that prospect, it would be worth collecting a large number of such parallel edges
in the latent space and exploring how the underlying proofs are related,
typically by inspecting their expansion trees,
to see whether our conjecture holds.
We leave that for future work.

\subsection{Contrarian edges and hard implications}
After observing the mainstream orientation of implication flows,
we looked for atomic edges with a minority or contrarian orientation.
To that purpose, we searched and collected a number of \emph{longest paths}
in the implication graph~$(G,\ImplicationOneStep)$ modulo~$\sim$.
Among these paths, we observed some unexpected behaviours
like \emph{swirls around the cliques or clusters} of provably equivalent theories.
A possible explanation is that these clusters of provably equivalent theories
are \emph{active proof-theoretic areas} of the latent space, where many different equivalent formulations
of the {very same concept} can be established by back-and-forth:
of the form:
\vspace{-.1em}
\begin{center}
$\Eqn{j}\Longrightarrow \Eqn{k}\Longrightarrow\Eqn{j}$
\end{center}
\vspace{-.1em}
%
as we observed empirically.

We also considered the hard implications mentioned on \href{https://teorth.github.io/equational_theories/blueprint/hard-implications-chapter.html}{Section 27 of the ET project blueprint}
as well as the implications with long Vampire times.
%
%
%
We noticed empirically that a number of them
had unusual and contrarian shapes as atomic edges in the latent space.
At the same time, we are not entirely sure of the phenomenon 
at this stage, and we could not develop a sufficiently clear and complete picture of it.
We thus prefer to leave this important and promising line of investigation for future work.




%


\section{Related works}\label{section/related-works}
%
%
%
An important question underlying our work is whether (and how) 
the latent information provided by finite model theory 
is related to the proof-theoretic
structure witnessed by the implication graph.
%
From that point of view, it would be instructive
%
%
to experiment with the convolutional neural network (CNN)
model developed in the ET project  \cite{Bolan2024equational}
and see whether it can be improved by integrating
the latent space structure coming from Stone pairings.

%

Our exploration of the connection between proof theory and finite model theory
is driven by the desire to understand the latent, statistical and distributional 
structure of deduction in machine learning,
and we hope that it will bring new insights on the way large language models 
reason internally in their own latent space~\cite{AspertiNaiboSacerdoti2025,zhou2025geometry}.
%
From that point of view, it is very much in line and inspired by the dualistic approach
to type theory and language developed in~\cite{GastaldiPellissier21,GastaldiPellissier23},
see also the more recent~\cite{BradleyGastaldiTerilla24}.

%

%







\section{Conclusion and future work}\label{section/conclusion}
This work started as a small-scale exercise 
in experimental logic, building on the 
achievements of the \href{https://teorth.github.io/equational_theories}{Equational Theories project}
initiated by Terence Tao fifteen months ago,
and on the exhaustive description
of the implication preorder~$\mathfont{(G,\ImplicationOrder)}$
between $4694$ equational theories established 
by this remarkably active and lively collaborative project.

Our plan was to proceed in two stages.
In a first stage, we would define a latent space 
of the $4694$ equational theories
using ideas and methods coming from machine learning and finite model theory.
Then, in a second stage, we would study the geometric (and metric) properties of the implication preorder $(G,\ImplicationOrder)$ of equational theories seen as \emph{embedded} in the latent space just constructed.
%

The project was propelled since its origins
by curiosity and exploratory joy, combined with the hope and desire of
revealing meaningful hidden latent structures at the heart of mathematical reasoning.
Despite the preliminary nature
and technical simplicity of this work,
the outcome of the experiment largely exceeds
what we originally expected:
%

\medbreak
\noindent
1. First, we observed with great surprise that the latent space was excessively good at detecting the signature of an equational theory, see \S\ref{section/latent-space-of-equational-theories}.
\medbreak
\noindent
2. Then, we measured significant differences between the length of an edge associated in the latent space
to a reversible proof and to a strict atomic (non reversible) proof.
From this follows that provably equivalent theories
are tightly clustered together in the latent space, see \S\ref{section/clustering-or-provably-equivalent-theories}.
\medbreak
\noindent
3. Finally, we discovered empirically the
existence of oriented \emph{implication flows} 
in the latent space of equational theories,
and studied the shape of hard implications, see \S\ref{section/implication-flows}.
%

\medbreak

This series of converging discoveries indicates an unexpected similitude
or correspondence between the purely \emph{model-theoretic} definition of the latent space based on Stone duality, and the purely \emph{proof-theoretic} definition of the implication graph between equational theories.
We believe that we only scratched the surface of a statistical variant (and refinement) of the celebrated G{\"o}del completeness theorem for first-order logic, see \cite{VanHeijenoort1967}.
We include in the Appendix a discussion on Herbrand theorem, which we believe should play a central role in that statistical reconstruction.

These empirical results also convey the hope of
enlarging the perspective beyond algebraic reasoning,
and defining 
a \emph{latent space of mathematical concepts}
building on the idea of a \emph{universe of types}
formulated by Martin-L{\"o}f \cite{MartinLof73}
and recently recast as 
a \emph{moduli space 
of types} in Voevodsky's homotopy type theory~\cite{hottbook}.

Finally, in a dual and complementary direction, our construction of the latent space
of equational theories could provide new tools and insights
in the search using reinforcement learning (or other methods)
for finite magmas satisfying specific algebraic laws, 
such as semigroups~\cite{Simpson21}.


%

\newpage

\section{Limitations}\label{section/limitations}
We introduce the latent space of equational theories and describe its main empiric properties. This is the first paper on the topic, and we thus open a new research area at the interplay of logic and machine learning. For that reason, much remains to be done.
%
The construction of the feature space is simplistic
and could be refined by taking random samples of tuples
in much larger finite magmas, or on a stochastic 
and non-deterministic extension of the usual notion of magma.
%
We also used a very elementary linear form of dimensionality reduction
based on a PCA method.
We could explore other more advanced dimensionality reduction methods such as $t$-SNE, see \cite{JMLR:v9:vandermaaten08a}.
%
%
Although this is mentioned for future work, the current paper
does not include any explicit and formal measure of complexity of proofs,
such as expansion trees coming from Herbrand theorem.
All these limitations can be seen as opening opportunities for future works.

\subsubsection*{Acknowledgments}
We are grateful to Mario Carneiro for the nice and inspiring talk
he gave on the Equational Theory Project at the AITP workshop in Aussois,
in September 2025.
This work has received funding from the European Research Council
under the European Union’s Horizon 2020 research and innovation
programme (Synergy Project Malinca, ERC Grant Agreement No 670624).

\bibliography{latent-reasoning.bib}



\appendix

\section{Linear regression graphs for the first\\
and second principal components~$X$ and~$Y$}
\begin{figure}[h!]
    \centering
    \includegraphics[width=0.9\linewidth]{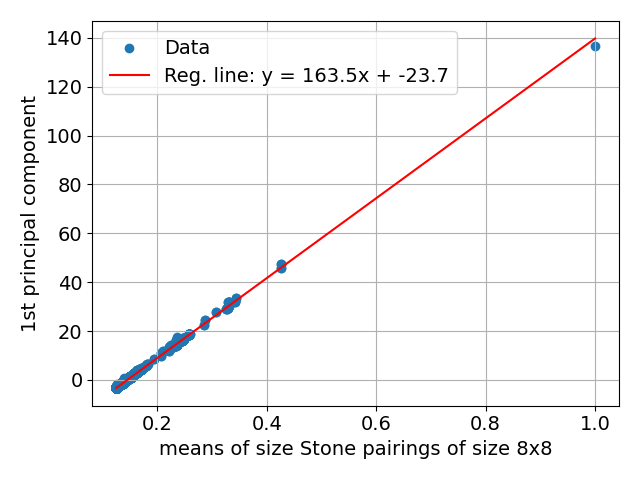}
    \caption{Linear regression graph of the expectation (or mean) of the Stone pairing for each equational theory with relation to the first principal component~$X$.}
    \label{fig:mean-pca}
\end{figure}

\begin{figure}
    \centering
    \includegraphics[width=0.9\linewidth]{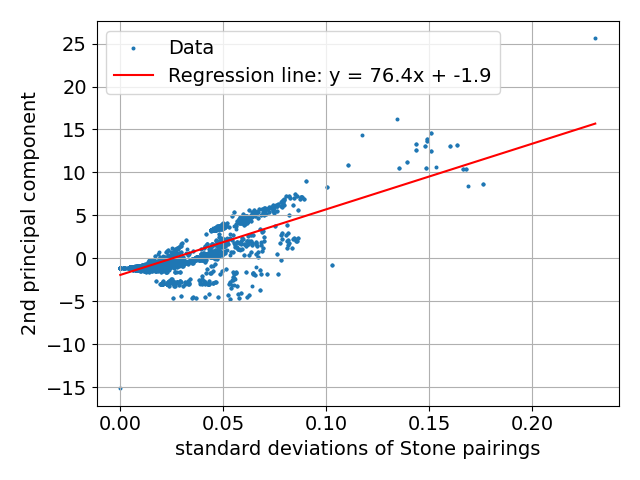}
    \caption{Regression graph of the standard value of the Stone pairings for each equation vrs. the second principal component~$Y$.}
    \label{fig:std-pca}
\end{figure}

We provide in Fig.~\ref{fig:mean-pca} and Fig.~\ref{fig:std-pca}
the linear regression graphs mentioned in \S\ref{subsection-X-axis} and in \S\ref{subsection-Y-axis}.

\section{Cliques of provably equivalent theories}

\begin{sidewaystable}
    \centering\tiny
\begin{tabular}{lrrrrrrrrrrrrrrrrrrrrrr}
\toprule
 & 1 & 2 & 3 & 4 & 5 & 6 & 7 & 8 & 9 & 12 & 14 & 18 & 27 & 30 & 31 & 37 & 40 & 71 & 76 & 112 & 419 & 1496 \\
\midrule
1 & 2717 & 312 & 102 & 36 & 80 & 48 & 0 & 64 & 108 & 0 & 0 & 0 & 0 & 0 & 0 & 0 & 0 & 0 & 0 & 0 & 0 & 0 \\
2 & 1174 & 164 & 156 & 144 & 0 & 48 & 0 & 0 & 0 & 0 & 0 & 0 & 0 & 0 & 248 & 0 & 0 & 0 & 0 & 0 & 0 & 0 \\
3 & 813 & 378 & 225 & 48 & 30 & 36 & 0 & 0 & 0 & 0 & 0 & 0 & 0 & 0 & 0 & 0 & 0 & 0 & 0 & 0 & 0 & 0 \\
4 & 328 & 64 & 24 & 32 & 0 & 48 & 0 & 0 & 0 & 0 & 0 & 0 & 0 & 0 & 0 & 0 & 0 & 0 & 0 & 0 & 0 & 0 \\
5 & 410 & 200 & 120 & 0 & 0 & 0 & 0 & 0 & 0 & 0 & 0 & 0 & 0 & 0 & 0 & 0 & 0 & 0 & 0 & 0 & 0 & 0 \\
6 & 606 & 228 & 234 & 96 & 0 & 216 & 0 & 96 & 0 & 72 & 0 & 0 & 0 & 180 & 0 & 0 & 0 & 0 & 0 & 0 & 0 & 0 \\
7 & 70 & 252 & 84 & 0 & 0 & 0 & 0 & 0 & 0 & 0 & 0 & 0 & 0 & 0 & 0 & 0 & 0 & 0 & 0 & 0 & 0 & 0 \\
8 & 512 & 80 & 240 & 0 & 0 & 0 & 0 & 64 & 0 & 0 & 0 & 0 & 0 & 0 & 0 & 0 & 0 & 0 & 0 & 0 & 0 & 0 \\
9 & 486 & 216 & 54 & 0 & 0 & 0 & 0 & 0 & 0 & 0 & 0 & 0 & 0 & 0 & 0 & 0 & 0 & 0 & 0 & 0 & 0 & 0 \\
12 & 252 & 72 & 0 & 0 & 0 & 0 & 0 & 0 & 0 & 0 & 0 & 0 & 0 & 0 & 0 & 0 & 0 & 0 & 0 & 0 & 0 & 0 \\
14 & 14 & 56 & 0 & 0 & 350 & 0 & 196 & 0 & 0 & 0 & 0 & 0 & 0 & 0 & 0 & 0 & 0 & 0 & 0 & 0 & 0 & 0 \\
18 & 72 & 0 & 108 & 144 & 0 & 0 & 0 & 0 & 0 & 0 & 0 & 0 & 0 & 0 & 0 & 0 & 1440 & 0 & 0 & 0 & 0 & 0 \\
27 & 216 & 0 & 324 & 0 & 0 & 0 & 0 & 0 & 0 & 0 & 0 & 0 & 0 & 0 & 0 & 0 & 2160 & 0 & 0 & 0 & 0 & 0 \\
30 & 510 & 300 & 0 & 0 & 0 & 0 & 0 & 480 & 0 & 0 & 0 & 0 & 0 & 0 & 0 & 0 & 0 & 0 & 0 & 0 & 0 & 0 \\
31 & 62 & 372 & 372 & 0 & 0 & 0 & 0 & 992 & 0 & 0 & 0 & 0 & 0 & 0 & 0 & 0 & 0 & 0 & 0 & 0 & 0 & 0 \\
37 & 370 & 296 & 0 & 0 & 0 & 0 & 0 & 0 & 0 & 0 & 0 & 1332 & 0 & 0 & 0 & 0 & 0 & 0 & 0 & 0 & 0 & 0 \\
40 & 240 & 320 & 0 & 0 & 0 & 0 & 0 & 640 & 0 & 0 & 0 & 0 & 0 & 0 & 0 & 0 & 0 & 0 & 0 & 0 & 0 & 0 \\
71 & 1704 & 0 & 426 & 568 & 0 & 0 & 0 & 0 & 0 & 0 & 0 & 0 & 0 & 0 & 0 & 5254 & 0 & 0 & 10792 & 15904 & 0 & 0 \\
76 & 760 & 608 & 912 & 608 & 0 & 0 & 0 & 0 & 0 & 0 & 0 & 0 & 0 & 0 & 0 & 0 & 6080 & 0 & 0 & 0 & 0 & 0 \\
112 & 2016 & 1792 & 672 & 0 & 0 & 1344 & 1568 & 0 & 2016 & 0 & 0 & 4032 & 6048 & 0 & 0 & 0 & 0 & 0 & 0 & 0 & 0 & 0 \\
419 & 4609 & 1676 & 1257 & 0 & 0 & 2514 & 0 & 0 & 0 & 0 & 5866 & 0 & 0 & 0 & 0 & 31006 & 0 & 0 & 0 & 0 & 0 & 0 \\
1496 & 59840 & 0 & 4488 & 5984 & 0 & 8976 & 0 & 0 & 0 & 0 & 0 & 0 & 0 & 0 & 0 & 0 & 0 & 212432 & 0 & 0 & 626824 & 0 \\
\bottomrule
\end{tabular}
\label{tab:ores}
\caption{Depicts the number of strict edges from each reversible edge to another.}
\end{sidewaystable}

The repartition of reversible cliques follows:

\medbreak
\begin{center}
\begin{footnotesize}
\begin{tabular}{|c|c|c|}
\hline
cardinal & number of cliques & number 
\\
of the clique & of this cardinal & of vertices
\\
\hline
1496 & 1 & 1496
\\
419 & 1 & 419
\\
112 & 2 & 224
\\
76 & 2 & 152
\\
71 & 2 & 142
\\
40 & 2 & 80
\\
37 & 2 & 74
\\
31 & 2 & 62
\\
30 & 1 & 30
\\
27 & 2 & 54
\\
18 & 2 & 36
\\
14 & 1 & 14
\\
12 & 1 & 12
\\
9 & 4 & 36
\\
8 & 7 & 56
\\
7 & 4 & 28
\\
6 & 11 & 66
\\
5 & 11 & 55
\\
4 & 14 & 56
\\
3 & 61 & 183
\\
2 & 137 & 274
\\
1 & 1145 & 1145
\\
\hline
\end{tabular}
\end{footnotesize}
\end{center}

\begin{figure}
\begin{center}
\begin{footnotesize}
\begin{tabular}{|c|l|}
\hline
size of & selected equation
\\
clique & in the clique
\\
\hline
$1496$ & $\Eqn{1}$ \hspace{5em} $x=y$
\\
$419$ & $\Eqn{45}$ \hspace{3em} $x\diamond y = z \diamond w$
\\
$112$ & $\Eqn{18}$ \hspace{3em} $x = y \diamond (z \diamond x)$
\\
$112$ & $\Eqn{26}$ \hspace{3em} $x = (x \diamond y) \diamond z$
\\
$76$ & $\Eqn{93}$ \hspace{2em} $x=y\diamond (z\diamond (w\diamond x))$
\\
$76$ & $\Eqn{268}$ \hspace{1.5em} $x=((x\diamond y)\diamond z)\diamond w$
\\
$71$ & $\Eqn{3}$ \hspace{4em} $x=x\diamond y$
\\
$71$ & $\Eqn{4}$ \hspace{4em} $x=y\diamond x$
\\
$40$ & $\Eqn{41}$ \hspace{3em} $x\diamond y = x \diamond z$
\\
$40$ & $\Eqn{44}$ \hspace{3em} $x\diamond y = z \diamond y$
\\
$37$ & $\Eqn{354}$ \hspace{2em} $x\diamond y = z\diamond (w\diamond y)$
\\
$37$ & $\Eqn{382}$ \hspace{2em} $x\diamond y = (x\diamond z)\diamond w$
\\
$31$ & $\Eqn{4378}$ \hspace{.5em} $x\diamond (y\diamond z) = w\diamond (u\diamond v)$
\\
$31$ & $\Eqn{4693}$ \hspace{.5em} $(x\diamond y)\diamond z = (w\diamond u)\diamond v$
\\
$30$
& $\Eqn{710}$ \hspace{1em} $x=y\diamond(y\diamond((x\diamond z)\diamond z))$
\\
& $\Eqn{2943}$ \hspace{.5em} $x=((y \diamond (y \diamond x))\diamond z)\diamond z$
\\
$27$ & $\Eqn{607}$ \quad $x = y\diamond (z\diamond (w\diamond (u\diamond x)))$
\\
$27$ & $\Eqn{3100}$ \quad $x = (((x\diamond y)\diamond z)\diamond w)\diamond u$
\\
$18$ & $\Eqn{3450}$ \quad $x\diamond y = z \diamond (w\diamond (u \diamond y))$
\\
$18$ & $\Eqn{4152}$ \quad $x\diamond y = ((x\diamond z)\diamond w)\diamond u$
\\
$14$ &  $\Eqn{4581}$ \quad $x\diamond (y \diamond z) = (w\diamond u)\diamond v$
\\
$12$ & $\Eqn{1354}$ \hspace{1em} $x=y\diamond (((z \diamond x)\diamond y) \diamond z)$
\\
& $\Eqn{2369}$ \hspace{1em} $x=(y\diamond (z \diamond (x\diamond y)))\diamond z$
\\
$9$ & $\Eqn{8}$ \hspace{3em} $x=x\diamond(x\diamond y) $
\\
$9$ & $\Eqn{27}$ \hspace{2.5em} $x=(y\diamond x)\diamond x$
\\
$9$ & $\Eqn{1815}$ \quad $x=(y\diamond z) \diamond ((w\diamond z)\diamond x))$
\\
$9$ & $\Eqn{1873}$ \quad $x= (x \diamond (y \diamond z )) \diamond (y\diamond w)$
\\
$8$ & $\Eqn{13}$ \hspace{2em} $x=y\diamond(x\diamond y)$
\\
& $\Eqn{28}$ \hspace{2em} $x=(y\diamond x)\diamond y$
\\
$8$ & $\Eqn{745}$ \quad $x = y\diamond (z\diamond (x\diamond y)\diamond z)$ 
\\
& $\Eqn{2978}$ \quad $x = ((y\diamond(z\diamond x))\diamond y)\diamond z$
\\
$8$ & $\Eqn{3369}$ \quad $x\diamond y = y\diamond (z\diamond (z\diamond x))$
\\
& $\Eqn{4181}$ \quad $x\diamond y = ((y\diamond z)\diamond z)\diamond x$
\\
$8$ & $\Eqn{4360}$ \hspace{1em} $x\diamond(y \diamond z) = x\diamond (w \diamond u)$
\\
$8$ & $\Eqn{4692}$ \hspace{1em} $(x\diamond y) \diamond z = (w \diamond u)\diamond z$
\\
$8$ & $\Eqn{4377}$ \hspace{1em} $x \diamond ( y \diamond z ) = w \diamond (u \diamond z)$
\\
$8$ & $\Eqn{4675}$ \hspace{1em} $(x\diamond y) \diamond z = (x \diamond w)\diamond u$
\\
$7$ &  $\Eqn{160}$ \hspace{2em} $x=(x\diamond y)\diamond (y\diamond z)$
\\
$7$
& $\Eqn{193}$ \hspace{2em} $x=(y\diamond z)\diamond (z\diamond x)$
\\
$7$ & $\Eqn{4467}$ \hspace{1em} $x\diamond(y\diamond x)=(z\diamond w)\diamond u$
\\
$7$ & $\Eqn{4579}$ \hspace{1em} $x\diamond(y\diamond z)=(w\diamond u)\diamond w$
\\
\hline
\end{tabular}
\end{footnotesize}
\end{center}
\caption{List of all equivalence classes of equational theories modulo~$\sim$,
of size between~$7$ and~$1496$, together with one or two distinctive elements of each clique.}
\end{figure}

\section{Herbrand theorem and parallelism of implication flows}\label{section/Herbrand-theorem}
%
%
%
%
Suppose given a pair of magma equations
\begin{center}
$\varphi(\mathbf{u})=\varphi(u_1,\dots,u_k)$ 
\end{center}
where $\varphi$ has free variables
$\mathbf{u}=(u_1,\dots,u_k)$ and
\begin{center}
$\psi(\mathbf{x})=\psi(x_1,\dots,x_\ell)$.
\end{center}
where $\psi$ has free variables $\mathbf{x}=(x_1,\dots,x_\ell)$.

\medbreak
A Herbrand proof
$$
(\theta_1, \dots, \theta_n) \quad : \quad 
\forall\mathbf{u}. \varphi(\mathbf{u}) \,\, \vdash \,\, \forall\mathbf{x}. \psi(\mathbf{x})
$$
is defined as a finite sequence
$(\theta_1, \dots, \theta_n)$
where each~$\theta_i$ is a substitution of the free variables
$$\mathbf{u}=(u_1,\dots,u_k)$$
of the equation~$\varphi$
with magma expressions parametrized by the free variables 
$$\mathbf{x}=(x_1,\dots,x_{\ell})$$
of the equation~$\psi$, such that one can establish
$$
\forall \mathbf{x}.
\hspace{.5em} \Big( \,\, \varphi\subst{\theta_1} \wedge \dots \wedge \varphi\subst{\theta_n} \,\, \Big)
\hspace{.5em} \Rightarrow \hspace{.5em} \phi(\mathbf{x}).
$$
Consider for instance $\Eqn{13}$
$$
\forall\mathbf{u}. \,\varphi(\mathbf{u}) \quad \equiv \quad \forall u. v. \,\,\, u = v \diamond (u \diamond v)
$$
and $\Eqn{1557}$
$$
\forall\mathbf{x}. \,\psi(\mathbf{x}) \hspace{.5em} \equiv \hspace{.5em}
\forall x. y. z. \,\,\,  x = (y \diamond z)\diamond (x \diamond (y \diamond z))
$$
with free variables~$\mathbf{u}=(u,v)$ and $\mathbf{x}=(x,y,z)$,
respectively.
In one direction, the fact that $\Eqn{13}$ implies $\Eqn{1557}$ is obvious 
by substituting the two variables $u$ and $v$ with the expressions 
\begin{center}
$\theta: u \mapsto x$ \quad and \quad $\theta : v \mapsto x \diamond y$.
\end{center}
From this, one concludes that the substitution~$\theta$
establishes the implication:
$$
\theta \quad : \quad 
\Eqn{13} \,\, \vdash \,\, \Eqn{1557} 
$$
Conversely, somewhat surprisingly,
the \href{https://teorth.github.io/equational_theories}{Equational Theories} project
indicates that $\Eqn{1557}$ implies $\Eqn{13}$.
This nontrivial fact can be established by applying the pair of substitutions
\begin{center}
$\theta_1 \hspace{.5em} : \hspace{.5em} x \mapsto u,$ \hspace{.5em}  $y\mapsto u \diamond v,$ 
\hspace{.5em}  $z \mapsto v \diamond (u\diamond u)$
\end{center}
\begin{center}
$\theta_2 \hspace{.5em} : \hspace{.5em} x \mapsto v,$  \hspace{.5em} $y \mapsto u,$ 
\hspace{.5em}   $z \mapsto u$.
\end{center}
to the equation
\begin{center}
$\psi(x,y,z) \hspace{1em} \equiv \hspace{1em}
x \,\, = \,\, (y \diamond z)\diamond (x \diamond (y \diamond z))$
\end{center}
in order to obtain the equation $\psi\subst{\theta_1}$
after substitution by~$\theta_1$:
\begin{center}
$u = ((u \diamond u)\diamond (v \diamond (u\diamond u)))\diamond (u \diamond ((u \diamond u)\diamond (v \diamond (u\diamond u))))$
\end{center}
and the equation $\psi\subst{\theta_2}$ after substitution by~$\theta_2$:
\begin{center}
$v = (u \diamond u)\diamond (v \diamond (u \diamond u))$
\end{center}
An easy equational reasoning establishes that the pair
of the two equations implies $\Eqn{13}$:
\begin{equation}\label{equation-13}
\varphi(u,v) \hspace{1em} \equiv \hspace{1em} u = v\diamond (u \diamond v)
\end{equation}
From this, one concludes that
$$
(\theta_1, \theta_2) \quad : \quad 
\Eqn{1557} \,\, \vdash \,\, \Eqn{13} 
$$
Note that the two substitutions~$\theta_3$ and~$\theta_4$
defined below would establish~(\ref{equation-13}) in just the same way:
\begin{center}
$\theta_3 \hspace{.5em} : \hspace{.5em} x \mapsto u,$ \hspace{.5em}  $y\mapsto u \diamond v,$ 
\hspace{.5em}  $z \mapsto v \diamond (u\diamond u)$
\end{center}
\begin{center}
$\theta_4 \hspace{.5em} : \hspace{.5em} x \mapsto v,$  \hspace{.5em} $y \mapsto u,$ 
\hspace{.5em}   $z \mapsto v$.
\end{center}
From this follows that
$$
(\theta_3, \theta_4) \quad : \quad 
\Eqn{1557} \,\, \vdash \,\, \Eqn{13} 
$$
This defines a category with equations as objects
and Herbrand proofs as morphisms.
The fact that $(\theta_1,\theta_2)\neq(\theta_3,\theta_4)$
means that this is a category and not just a preorder.

There are special commutative triangles:
\begin{center}
\begin{small}
\begin{tikzcd}[column sep = 1.2em, row sep = .8em]
&&
{\forall \mathbf{u}. \psi(\mathbf{u})}
\arrow[dd,"\alpha"]  
\\
{\forall \mathbf{x}. \phi(\mathbf{x})}
\arrow[rru,"\theta"]
\arrow[rrd,"\alpha\circ\theta"{swap}]
\\
&& {\forall \mathbf{v}. \psi(\mathbf{v})}
\end{tikzcd}
\begin{tikzcd}[column sep = 1.2em, row sep = .8em]
{\forall \mathbf{x}. \phi(\mathbf{x})}\arrow[rrd,"\theta\circ\alpha"]\arrow[dd,"\alpha"{swap}]  
\\
&& {\forall \mathbf{u}. \psi(\mathbf{u})}
\\
{\forall \mathbf{y}. \phi(\mathbf{y})}\arrow[rru,"\theta"{swap}] &&
\end{tikzcd}
\end{small}
\end{center}
where the sides are linear substitutions.
These deserve to be filled, in order to define a higher-dimensional manifold 
(defined as a simplicial set) of equations and reasonings,
constructed as a variation of the nerve of a category.
We leave that for future work.

\end{document}